\documentclass[twocolumn,journal]{IEEEtran}
\usepackage[LGR,T1]{fontenc}
\usepackage{color}
\usepackage{float}
\usepackage{amsmath}
\usepackage{amsthm}
\usepackage{amssymb}
\usepackage{graphicx}
\usepackage{setspace}

\makeatletter

\DeclareRobustCommand{\greektext}{%
  \fontencoding{LGR}\selectfont\def\encodingdefault{LGR}}
\DeclareRobustCommand{\textgreek}[1]{\leavevmode{\greektext #1}}
\ProvideTextCommand{\~}{LGR}[1]{\char126#1}

\floatstyle{ruled}
\newfloat{algorithm}{tbp}{loa}
\providecommand{\algorithmname}{Algorithm}
\floatname{algorithm}{\protect\algorithmname}

\theoremstyle{plain}
\newtheorem{thm}{\protect\theoremname}
\theoremstyle{plain}
\newtheorem{lem}[thm]{\protect\lemmaname}
\theoremstyle{plain}
\newtheorem{prop}[thm]{\protect\propositionname}

\usepackage[caption=false,font=footnotesize]{subfig}
\usepackage{algorithm,algpseudocode}
\allowdisplaybreaks

\makeatother

\providecommand{\lemmaname}{Lemma}
\providecommand{\propositionname}{Proposition}
\providecommand{\theoremname}{Theorem}

\begin{document}

\title{Low Complexity Resource Allocation for \\Massive Carrier Aggregation}

\author{Stelios~Stefanatos, Fotis~Foukalas, \IEEEmembership{Member,~IEEE},
and Theodoros A. Tsiftsis, \IEEEmembership{Senior Member,~IEEE}\thanks{S. Stefanatos is with the Industrial Systems Institute, Athena Research
and Innovation Centre, Patras, Greece (e-mail: sstefanatos@isi.gr).}\thanks{F. Foukalas is with the Industrial Systems Institute, Athena Research
and Innovation Centre, Patras, Greece (e-mail: foukalas@isi.gr).}\thanks{T. A. Tsiftsis is with the School of Engineering, Nazarbayev University,
Astana 010000, Kazakhstan (e-mail: theodoros.tsiftsis@nu.edu.kz).}\thanks{This work has been supported by the \textquotedblright Spectrum Overlay
through Aggregation of Heterogeneous Dispersed Bands\textquotedblright{}
project, ICT-SOLDER, www.ict-solder.eu, FP7 grant agreement number
619687.}}
\maketitle
\begin{abstract}
Optimal resource allocation (RA) in massive carrier aggregation scenarios
is a challenging combinatorial optimization problem whose dimension
is proportional to the number of users, component carriers (CCs),
and OFDMA resource blocks per CC. Towards scalable, near-optimal RA
in massive CA settings, an iterative RA algorithm is proposed for
joint assignment of CCs and OFDMA resource blocks to users. The algorithm
is based on the principle of successive geometric programming approximations
and has a complexity that scales only linearly with the problem dimension.
Although its derivation is based on a relaxed formulation of the RA
problem, the algorithm is shown to converge to integer-valued RA variables
with probability 1 under mild assumptions on the distribution of user
utilities. Simulations demonstrate improved performance of the proposed
algorithm compared to commonly considered heuristic RA procedures
of comparable complexity.
\end{abstract}


\section{Introduction}

\IEEEPARstart{C}{arrier} aggregation (CA) is considered as one of
the key features of future cellular networks for effectively increasing
the system bandwidth by simultaneous utilization of multiple component
carriers (CCs) \cite{CA mag}. Although up to $5$ CCs were initially
considered for CA in the Long-Term-Evolution-Advanced (LTE-A) standard,
the increasing system demands strongly suggest that future applications
of CA will utilize more CCs, towards realizing the concept of \emph{massive}
CA \cite{massive CA mag}. Proposals for CA of $32$ CCs in LTE have
appeared \cite{32 CA}.

One of the major challenges of CA is the resource allocation (RA).
With each CC typically consisting of multiple OFDMA resource blocks
(RBs) and with limitations on the maximum number of CCs that UEs can
utilize for communication, \emph{optimal} RA becomes a constrained,
integer optimization problem that is difficult to solve, even for
a small number of available CCs \cite{RRM survey,greedy algorithm}.
This has led to most RA proposals performing CC and RB allocation
separately \cite{RRM survey,Mogensen}, with CC allocation performed
first by some heuristic method, followed by RB allocation per CC.
Towards improving system performance, the problem of \emph{joint}
CC and RB allocation was considered in \cite{greedy algorithm,WSCP},
where iterative RA algorithms were proposed in an attempt to reduce
the number of combinations required to be examined by a brute-force
solution approach. However, the complexity of these algorithms becomes
impractical when a massive CA scenario is considered.

In this correspondence, the problem of joint CC/RB allocation to UEs
in a massive CA scenario (e.g., with $50$ available CCs) is considered,
with the goal of maximizing the weighted sum utility of the users.
Towards obtaining an efficient, scalable RA algorithm, the method
of successive geometric programming approximations \cite{PC by GP}
is applied to a relaxed formulation of the RA problem. The resulting
iterative RA algorithm has a simple analytical representation (no
general-purpose numerical optimization procedures are required) with
its complexity scaling only linearly with the problem dimension when
the maximum number of iterations is kept fixed. It is shown that,
under mild assumptions on the distribution of utility functions, the
algorithm converges with probability $1$ to integer-valued RA variables,
which, in certain special cases of the RA problem, are the optimal
ones. Performance of the algorithm in a massive CA setting is investigated
numerically, where it is shown to outperform heuristic RA approaches
of comparable complexity.

\section{System Model and Problem Formulation}

The downlink or uplink of a single cell is considered that serves
$K\geq2$ user equipments (UEs), indexed by $k\in\mathcal{K}\triangleq\{1,2,\ldots,K\}$.
The system has $M\geq2$ available CCs with transmissions over the
$m$-th CC, $m\in\mathcal{M}\triangleq\{1,2,\ldots,M\}$, performed
via OFDMA. Without loss of generality, it will be assumed that all
CCs have the same bandwidth that is partitioned into $N\geq2$ RBs
of equal size, indexed by $n\in\mathcal{N}\triangleq\{1,2,\ldots,N\}$. 

Let $\phi_{k,m,n}>0$ denote the utility that the $k$-th UE achieves
when utilizing the $n$-th RB of the $m$-th CC. A commonly used metric
for resource allocation (RA) purposes is the weighted sum utility
($\mathsf{WSU}$) of UEs \cite{greedy algorithm,WSCP,Giannakis},
defined as

\begin{equation}
\mathsf{WSU}\triangleq\sum_{k\in\mathcal{K}}w_{k}\sum_{m\in\mathcal{M}}\sum_{n\in\mathcal{N}}\alpha_{k,m,n}\phi_{k,m,n},\label{eq:wsu_Def}
\end{equation}

\noindent where $w_{k}>0$ is the weight of UE $k$, and $\{\alpha_{k,m,n}\}$
is a set of $KMN$ binary-valued ($0-1$) RA variables reflecting
whether UE $k$ is allocated to RB $n$ of CC $m$ or not ($\alpha_{k,m,n}=1,0$,
respectively). With the goal of maximizing the $\mathsf{WSU}$ (equivalently,
minimizing $1/\mathsf{WSU}$), the RA variables are obtained as the
solution of the constrained integer (binary) optimization problem
described in (\ref{eq:RA problem original}), where $\mathcal{M}_{k}\triangleq\{m\in\mathcal{M}:\alpha_{k,m,n}=1\text{ for some }n\text{\ensuremath{\in\mathcal{N}}}\}$
is the set of CCs where UE $k$ is allocated at least one RB and $|\mathcal{S}|$
denotes the number of elements of the set $\mathcal{S}$.

\begin{equation}
\left\{ \begin{aligned}\text{minimize } & 1/\mathsf{WSU},\\
\text{subject to } & \mathsf{C1}:\sum_{k\in\mathcal{K}}\mathcal{\alpha}_{k,m,n}\leq1,\forall m,n,\\
 & \mathsf{C2}:|\mathcal{M}_{k}|\leq M_{k},\forall k,\\
 & \mathsf{C3}:|\cup_{k\in\mathcal{K}}\mathcal{M}_{k}|\leq M_{0},\\
 & \mathsf{C4}:\alpha_{k,m,n}\in\{0,1\},\forall k,m,n,
\end{aligned}
\right\} \label{eq:RA problem original}
\end{equation}

Constraint $\mathsf{C1}$ corresponds to the common requirement that
at most one UE is allocated to the $n$-th RB of the $m$-th CC. Constraint
$\mathsf{C2}$ restricts the number of CCs used by UE $k$ to a maximum
value $M_{k}\leq M$, possibly different among UEs. Constraint $\mathsf{C3}$
guarantees that $M_{0}\leq M$ CCs in total will be utilized by the
system for RA purposes. The last two constraints may be imposed in
practice due to, e.g., compatibility with legacy devices that can
communicate only via a single CC, power consumption considerations
when a UE operates on multiple CCs at the same time, and utilization
of the remaining $M-M_{0}$ CCs for other system applications of lower
priority.

For the special case where $M_{k}=M_{0}=M$ for all $k\in\mathcal{K}$,
constraints $\mathsf{C2}$ and $\mathsf{C3}$ become irrelevant and
the RA problem effectively corresponds to a standard OFDMA resource
allocation problem on a single CC with $MN$ RBs \cite{Giannakis}.
The optimal allocation in this case is a simple, ``winner-takes-all''
assignment per RB \cite{Giannakis}, namely, 
\begin{equation}
\alpha_{k^{*},m,n}=\begin{cases}
1, & k^{*}=\arg\underset{k\in\mathcal{K}}{\max}\:w_{k}\phi_{k,m,n},\\
0, & k\neq k^{*},
\end{cases}\label{eq:greedy_assigments}
\end{equation}
for all $m,n$, with ties resolved arbitrarily. This algorithm has
a complexity that scales only linearly with number of RA variables.
However, for the general case where limitations are imposed on the
maximum number of employed CCs, optimal RA requires solving a combinatorial
problem whose complexity scales exponentially with the number of RA
variables. For massive CA applications, optimal solution of the RA
allocation problem becomes impractical, which motivates the search
for alternative RA procedures.

\section{Low-Complexity Resource Allocation Algorithm}

Towards obtaining an efficient, low complexity RA algorithm for massive
CA scenarios, the original RA problem can be reformulated by introducing
two sets of \emph{auxiliary }RA variables, $\{\beta_{k,m}\}$, $\{\gamma_{m}\}$
of cardinality $KM$, and $M$, respectively. The RA variables $\{\beta_{k,m}\}$
indicate whether UE $k$ is allocated to CC $m$ or not ($\beta_{k,m}=1,0$,
respectively) and the RA variables $\{\gamma_{m}\}$ indicate whether
CC $m$ is utilized for transmissions by the system or not ($\gamma_{m}=1,0$,
respectively). By expressing the $\mathsf{WSU}$ in the equivalent
form 

\begin{equation}
\mathsf{WSU}=\sum_{k\in\mathcal{K}}w_{k}\sum_{m\in\mathcal{M}}\gamma_{m}\beta_{k,m}\sum_{n\in\mathcal{N}}\alpha_{k,m,n}\phi_{k,m,n},\label{eq:wsu_auxiliary}
\end{equation}

\noindent and treating $\{\alpha_{k,m,n}\}$, $\{\beta_{k,m}\}$,
$\{\gamma_{m}\}$ as independent optimization variables that are continuous-valued
in the interval $(0,1]$, a \emph{relaxed} version of the original
RA problem can be formulated as

\begin{equation}
\left\{ \begin{aligned}\text{minimize } & 1/\mathsf{WSU},\\
\text{subject to } & \mathsf{C1}:\sum_{k\in\mathcal{K}}\mathcal{\alpha}_{k,m,n}\leq1,\forall m,n,\\
 & \mathsf{C2'}:\sum_{m\in\mathcal{M}}\beta_{k,m}\leq M_{k},\forall k,\\
 & \mathsf{C3'}:\sum\gamma_{m}\leq M_{0}\\
 & \mathsf{C4'}:\alpha_{k,m,n},\beta_{k,m},\gamma_{m}\in(0,1],\forall k,m,n.
\end{aligned}
\right\} \label{eq:RA problem auxiliary}
\end{equation}

Note that the introduction of the auxiliary variables, although increasing
the dimension of the problem as a total of $KMN+KM+M$ variables have
to be found. However, it allows for expressing the combinatorial constraints
$\mathsf{C2}$ and $\mathsf{C3}$ of the original RA problem formulation
in the much more convenient, linear formulation of $\mathsf{C2}'$
and $\mathsf{C3}'$, respectively. Consideration of strictly positive
values for the RA variables (even though they can be equal to zero,
in principle) is only a technical requirement for the following algorithm
development and has no effect in practice, since a RA variable of
value less than a sufficiently small positive threshold may be safely
assumed as zero. 

Consideration of the relaxed RA problem is motivated by noting that
its formulation corresponds to that of a convex, geometric programming
(GP) problem \cite{Boyd}, with the exception that the objective function
is not a posynomial with respect to (w.r.t.) the RA variables, but
the inverse of a posynomial. This exact type of problem was considered
in \cite{PC by GP} where an iterative solution algorithm was proposed
based on solving a sequence of successive GP approximations of the
problem formulation. The algorithm is guaranteed to converge to a
point satisfying the Karush-Kuhn-Tucker (KKT) conditions of the problem. 

Applying the same procedure as in \cite{PC by GP} for solving the
relaxed RA problem results in obtaining the (positive-valued) estimates
$\{\alpha_{k,m,n}^{(i)}\},\{\beta_{k,m}^{(i)}\},\{\gamma_{m}^{(i)}\}$
of the RA variables at iteration $i$, as the solution of a modified
version of the relaxed RA problem where the objective function $1/\mathsf{WSU}$
is approximated by the monomial (see \cite{PC by GP} for details)

\begin{equation}
\tilde{f}^{(i)}=\prod_{k,m,n}\left(\frac{w_{k}\phi_{k,m,n}\beta_{k,m}\gamma_{m}\alpha_{k,m,n}}{u_{k,m,n}^{(i)}}\right)^{-u_{k,m,n}^{(i)}},\label{eq:f_tilde}
\end{equation}

\noindent where

\begin{equation}
u_{k,m,n}^{(i)}\triangleq\frac{w_{k}\phi_{k,m,n}\beta_{k,m}^{(i-1)}\gamma_{m}^{(i-1)}\alpha_{k,m,n}^{(i-1)}}{\sum_{k',m',n'}w_{k'}\phi_{k',m',n'}\beta_{k',m'}^{(i-1)}\gamma_{m'}^{(i-1)}\alpha_{k',m',n'}^{(i-1)}},\label{eq:u_def}
\end{equation}

\noindent for all $k,m,n$, where $\{\alpha_{k,m,n}^{(i-1)}\},\{\beta_{k,m}^{(i-1)}\},\{\gamma_{m}^{(i-1)}\}$
are the estimates of the RA variables obtained at iteration $i-1$.
As the following result shows, the new RA variables estimates can
be obtained via a simple (semi-) closed form formula, with no need
to employ general-purpose numerical solvers.
\begin{lem}
\label{lem:scalar iteration}The optimal RA variables for the GP optimization
problem resulting by replacing the objective function $1/\mathsf{WSU}$
in the relaxed RA problem formulation with $\tilde{f}^{(i)}$ as defined
in (\ref{eq:f_tilde}) equals
\end{lem}
\begin{eqnarray*}
\alpha_{k,m,n}^{(i)} & = & \alpha_{k,m,n}^{(i-1)}\frac{\beta_{k,m}^{(i-1)}w_{k}\phi_{k,m,n}}{\sum_{k'}\alpha_{k',m,n}^{(i-1)}\beta_{k',m}^{(i-1)}w_{k'}\phi_{k',m,n}}\\
\beta_{k,m}^{(i)} & = & \min\left\{ 1,\beta_{k,m}^{(i-1)}\frac{\sum_{n}w_{k}\gamma_{m}^{(i-1)}a_{k,m,n}^{(i-1)}\phi_{k,m,n}}{\lambda_{k}^{(i)}}\right\} ,\\
\gamma_{m}^{(i)} & = & \min\left\{ 1,\gamma_{m}^{(i-1)}\frac{\sum_{k,n}w_{k}\beta_{k,m}^{(i-1)}a_{k,m,n}^{(i-1)}\phi_{k,m,n}}{\mu^{(i)}}\right\} ,
\end{eqnarray*}
for all $k\in\mathcal{K},m\in\mathcal{M},n\in\mathcal{N},$ where
$\lambda_{k}^{(i)}>0$, $\mu^{(i)}>0$, are uniquely determined by
the conditions $\sum_{m}\beta_{k,m}^{(i)}=M_{k}$, $\sum_{m}\gamma_{m}^{(i)}=M_{0}$,
respectively.
\begin{IEEEproof}
Recalling that replacing the objective function of an optimization
problem with its logarithm does not change the solution of the optimization
variables \cite{Boyd}, the objective $\tilde{f}^{(i)}$ can be replaced
by $\log(\tilde{f}^{(i)})$, which is equal to

\[
-\sum_{k,m,n}u_{k,m,n}^{(i)}(\log(a_{k,m,n})+\log(\beta_{k,m})+\log(\gamma_{m})),
\]

\noindent after dropping additive constants that are independent of
the RA variables and play no role in the solution. It can be verified
that the resulting problem formulation is convex. Therefore, standard
solution techniques using Lagrange multipliers can be employed \cite{Boyd}
resulting in the optimal RA variables stated in the Lemma. Details
are omitted.
\end{IEEEproof}
The successive GP approximations (SGPA) RA algorithm is summarized
at the top of the page. Its complexity scales only linearly with the
number of RA variables, as long as the number of iterations required
for convergence is independent of the dimension of the problem. Even
though the latter condition is not the case, simulations show that
limiting the number of iterations to a maximum number, irrespective
of the dimension of the problem, yields good performance. 

\begin{algorithm}[t]
\begin{singlespace}
\noindent \begin{centering}
\caption{\label{tab:SGPA-RA-Algorithm}SGPA RA Algorithm}
\par\end{centering}
\end{singlespace}
\begin{enumerate}
\item Choose an initial estimate $a_{k,m,n}^{(0)}>0$, $\beta_{k,m}^{(0)}>0$,
$\gamma_{m}^{(0)}>0$, $\forall k,m,n$, of the RA variables.
\item At iteration $i\geq1$, obtain a new estimate of the RA variables
$\{a_{k,m,n}^{(i)}\},\{\beta_{k,m}^{(i)}\},\{\gamma_{m}^{(i)}\}$
as per Lemma \ref{lem:scalar iteration}.
\item Increase iteration index $i$ and go to step 2 until convergence to
a fixed point.
\end{enumerate}
\end{algorithm}

\section{Convergence Properties of the SGPA RA Algorithm}

Although the SGPA RA is guaranteed to converge to a KKT point of the
relaxed RA problem formulation, it is of interest to determine further
properties of the convergence in order to obtain insights on its operation
and usefulness of provided solutions. The key to study the convergence
is to view the parallel updates of the scalar RA variables performed
by the algorithm as parallel updates of appropriate vector RA variables.
In particular, consider (a) the RB allocation vector variables $\{\ensuremath{\boldsymbol{\alpha}_{m,n}}\}$,
where $\ensuremath{\boldsymbol{\alpha}_{m,n}}\triangleq[\alpha_{1,m,n},\alpha_{2,m,n},\ldots,\alpha_{K,m,n}]$,
(b) the UE-CC allocation vector variables $\{\ensuremath{\boldsymbol{\beta}_{k}}\}$,
where $\boldsymbol{\beta}_{k}\triangleq[\beta_{k,1},\beta_{k,2},\ldots,\beta_{k,M}]$,
and (c) the CC activation vector variable $\boldsymbol{\gamma}\triangleq[\gamma_{1},\gamma_{2},\ldots,\gamma_{M}]$.
Using the compact notation $\mathbf{x}=[x_{1},x_{2,}\ldots,x_{P}]$
for representing any of the aforementioned vector variables, and according
to Lemma \ref{lem:scalar iteration}, the update rules of the SGPA
RA algorithm are equivalent to updating the elements of $\mathbf{x}$
at iteration $i$ as 

\begin{equation}
x_{p}^{(i)}=\min\left\{ 1,x_{p}^{(i-1)}\frac{r_{p}^{(i)}}{\kappa^{(i)}}\right\} ,p\in\mathcal{P}\triangleq\{1,2,\ldots,P\},\label{eq:general_update_rule}
\end{equation}

\noindent where $r_{p}^{(i)}>0,p\in\mathcal{P}$, is independent of
$\mathbf{x}^{(i)}$ and $\kappa^{(i)}>0$ is selected such that 

\begin{equation}
\sum_{p=1}^{P}x_{p}^{(i)}=L,\text{ for all }i,\label{eq:sum constraint-1}
\end{equation}

\noindent where $L\in\mathcal{P}$. For example, with $\mathbf{x}$
corresponding to $\ensuremath{\boldsymbol{\beta}_{k}}$, for some
$k\in\mathcal{K}$, index $p$ in (\ref{eq:general_update_rule})
represents the tuple $(k,m)$ with $m\in\mathcal{M}$, $\mathcal{P}=\mathcal{M}$,
$P=M$, $\kappa^{(i)}$=$\lambda_{k}^{(i)}$, $L=M_{k}$, and $r_{p}^{(i)}=w_{k}\gamma_{m}^{(i-1)}\sum_{n}\alpha_{k,m,n}^{(i-1)}\phi_{k,m,n}$.

The following technical result is fundamental towards understanding
the operation of the SGPA RA algorithm and its convergence properties.
\begin{prop}
\label{prop:hardware arithmentic}Let the iterations of (\ref{eq:general_update_rule})
initialized by $\{x_{p}^{(0)}\}_{p\in\mathcal{P}},$ such that it
holds $x_{p}^{(0)}\in(0,1]$, for all $p\in\mathcal{P}$, and assume
that there exists a permutation $\pi$ of $\mathcal{P}$ and an integer
$P_{0}\in\mathcal{P}$ such that $r_{\pi(1)}^{(i)}>r_{\pi(2)}^{(i)}>\cdots>r_{\pi(P_{0})}^{(i)}>0$
and $r_{\pi(P_{0}+1)}^{(i)}=\cdots=r_{\pi(P)}^{(i)}=0$, for all $i\geq0$.
When the iterations are performed with finite (hardware) precision
arithmetic, sequence $x_{p}^{(i)},p\in\mathcal{P}$, converges to
the limit
\end{prop}
\begin{equation}
\bar{x}_{p}=\begin{cases}
1, & \text{if }r_{p}^{(0)}>0\text{ and }r_{p}^{(0)}\geq r_{\pi(L)}^{(0)},\\
0, & \text{if }r_{p}^{(0)}=0\text{ or }r_{p}^{(0)}<r_{\pi(L)}^{(0)},
\end{cases}.\label{eq:fixed point}
\end{equation}

\begin{IEEEproof}
See Appendix.
\end{IEEEproof}
\emph{Remark}: In case where $x_{p}^{(0)}=0$ for some $p\in\mathcal{P}$,
it can be verified that $x_{p}^{(i)}=0$ for all $i\geq0$, and the
iterations of (\ref{eq:general_update_rule}) are essentially performed
for the sequence of indices in $\mathcal{P}\setminus p$ with Prop.
\ref{prop:hardware arithmentic} changed accordingly. 

In order to obtain insights on the operation of the SGPA RA algorithm,
consider the special case where $\{\alpha_{k,m,n}\}$ and $\{\gamma_{m}\}$
have been pre-selected (by external means) and the optimal $\{\beta_{k,m}\}$
are required. It is easy to see that in this case, the RA problem
becomes a linear programming (LP) problem with a particularly simple
solution \cite{Boyd}: Assign to UE $k\in\mathcal{K}$ the $M_{k}$
CCs corresponding to the largest effective CC utilities $\bar{\phi}_{k,m},m\in\mathcal{M}$,
where $\bar{\phi}_{k,m}\triangleq w_{k}\gamma_{m}\sum_{n}\alpha_{k,m,n}\phi_{k,m,n}$.
Solving the same problem with the SGPA RA algorithm, results in the
iterative estimates of $\{\beta_{k,m}\}$ for UE $k$ corresponding
to the general update formula of (\ref{eq:general_update_rule}) with
$r_{k,m}^{(i)}=\bar{\phi}_{k,m}$ for all $i$, which, by Prop. \ref{prop:hardware arithmentic},
converge to the optimal solution if the positive elements of $\{\bar{\phi}_{k,m}\}_{m\in\mathcal{M}}$
are not equal. 

\begin{figure}[t]
\noindent \centering{}\includegraphics[width=0.9\columnwidth]{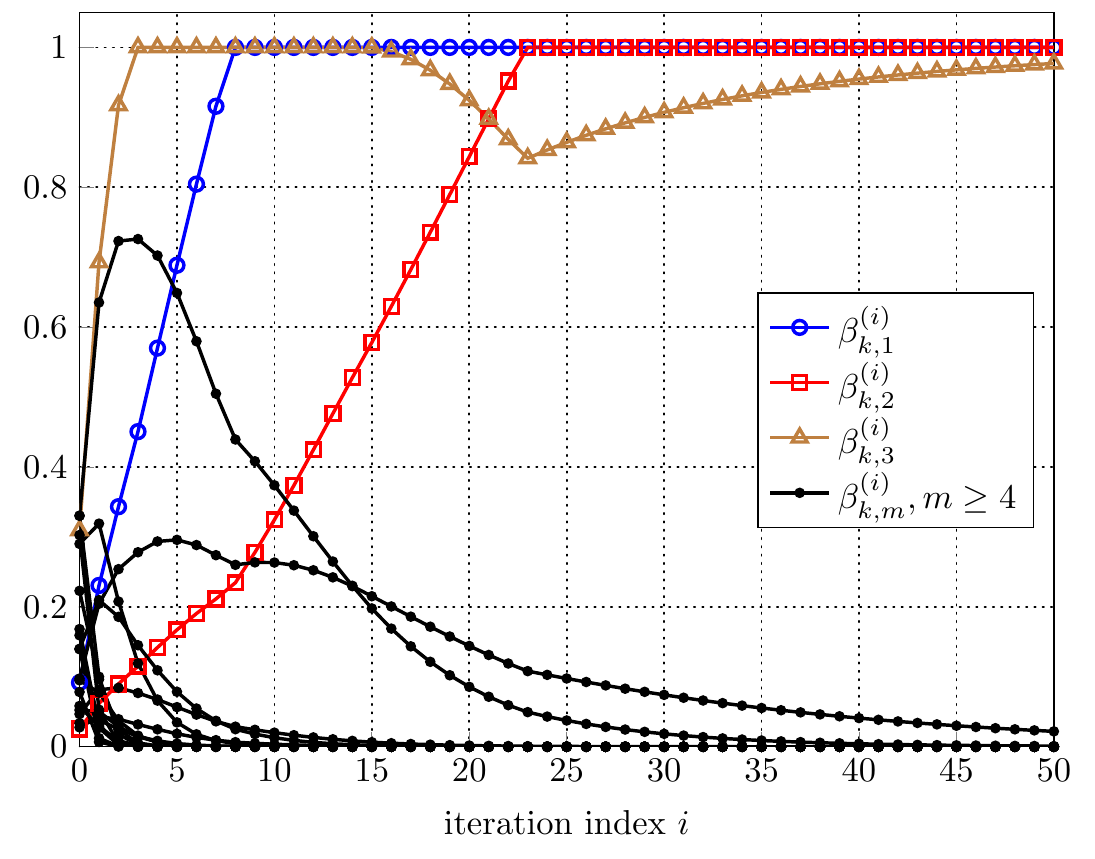}\caption{Convergence example of the SGPA RA algorithm w.r.t.\textcolor{red}{{}
}$\{\beta_{k,m}\}_{m\in\mathcal{M}}$ when $\{\alpha_{k,m,n}\}$ and
$\{\gamma_{m}\}$ are given ($M=20,M_{k}=3$).\label{fig:iterations}}
\end{figure}

A typical example of the convergence of the $\{\beta_{k,m}\}_{m\in\mathcal{M}}$
estimates for a UE $k$ obtained by the SGPA RA algorithm in this
special case is shown in Fig. \ref{fig:iterations}, with $M=20$
and $M_{k}=3$. The effective weights $\{\bar{\phi}_{k,m}\}_{m\in\mathcal{M}}$
were randomly generated and ordered such that $\bar{\phi}_{k,1}>\bar{\phi}_{k,2}>\ldots>\bar{\phi}_{k,M}>0$,
and a random initialization $\{\beta_{k,m}^{(0)}>0\}_{m\in\mathcal{M}}$
such that $\sum_{m\in\mathcal{M}}\beta_{k,m}=3$ was used. It can
be seen that the RA variables converge to $\beta_{k,m}=1$ for $m=1,2,3$,
as predicted by Prop. \ref{prop:hardware arithmentic}. Note that,
even though at $i=50$, $\beta_{k,3}$ has not achieved the value
of $1$, stoping the iterations at $i=15$ and quantizing the RA variables
to their nearest integer would still result in the optimal RA variables. 

Although the above result demonstrates that the SGPA RA algorithm
converges to the optimal RA variables under special cases, this is
not the case in general, i.e., a suboptimal convergence point is expected.
However, as the following result shows, the convergence point has
desirable properties, namely, it is binary-valued irrespective of
the initialization as long a mild condition on the utilities holds.
\begin{prop}
\label{prop:Global convergence}Under that assumption that $\{\phi_{k,m,n}\}$
are distributed according to a joint continuous probability density
function (p.d.f.) over $[0,\infty)^{KMN}$, and for any initialization
of the RA variables that is independent of $\{\phi_{k,m,n}\}$, the
SGPA RA algorithm converges with probability $1$ to binary-valued
RA variables.
\end{prop}
\begin{IEEEproof}
Since the SGP RA algorithm is guaranteed to converge to a KKT point
of the relaxed RA problem, it follows that that the sequences $\{r_{p}^{(i)}\}_{p\in\mathcal{P}}$
of the general iteration formula of (\ref{eq:general_update_rule})
will converge to a limit $\{\bar{r}_{p}\}_{p\in\mathcal{P}}$. Assuming
for the moment that no two positive elements of $\{\bar{r}_{p}\}_{p\in\mathcal{P}}$
are the same, it follows from fundamental properties of limits that
for any sufficiently small $\epsilon>0$, there exists an iteration
index $i_{\epsilon}$, a permutation $\pi$ of $\mathcal{P}$, and
an integer $P_{0}\in\mathcal{P}$ such that it holds $|r_{p}^{(i)}-\bar{r}_{p}|<\epsilon$
for all $p\in\mathcal{P}$, with $r_{\pi(1)}^{(i)}>r_{\pi(2)}^{(i)}>\cdots>r_{\pi(P_{0})}^{(i)}>0$
, $r_{\pi(P_{0}+1)}^{(i)}=\cdots=r_{\pi(P)}^{(i)}=0$, for all $i\geq i_{\epsilon}$.
That is, as the algorithm approaches one of its fixed points, the
elements of $\{r_{p}^{(i)}\}_{p\in\mathcal{P}}$, although varying
in principle as the iterations progress, will achieve an ordering
that holds for all $i\geq i_{\epsilon}$. Treating $\{x_{p}^{(i_{\epsilon})}\}$
as an initialization point for the algorithm iterations, it follows
from Prop. \ref{prop:hardware arithmentic} that each RA variable
will converge to either $0$ or $1$.

Next, it will be shown that, with probability one, the limit $\{\bar{r}_{p}\}_{p\in\mathcal{P}}$
has no two positive elements that are equal. By examination of the
equations corresponding to the KKT conditions of the relaxed RA problem,
it can be verified that the condition $\bar{r}_{p}=\bar{r}_{q}$,
for some $p\neq q$, does not hold identically (by default) at any
KKT point. Noting that both the equations corresponding to the KKT
conditions as well as the expressions for $r_{p}^{(i)}$ (and hence,
$\bar{r}_{p}$) are posynomials w.r.t. the RA variables and $\{\phi_{k,m,n}\}$,
it follows that the set $\mathcal{Z}\triangleq\{\{\phi_{k,m,n}\}:\bar{r}_{p}=\bar{r}_{q}>0,\text{ for some }p\neq q\}$
is of (Lebesgue) measure zero \cite[Corollary 10]{Cunning}, which
implies that the probability of the event $\{\phi_{k,m,n}\}\in\mathcal{Z}$
is zero. 
\end{IEEEproof}
\emph{Remark}: The assumption of continuous-valued $\{\phi_{k,m,n}\}$
is critical for the operation and convergence of the proposed algorithm
and applies to many possible utilities that can be considered for
RA purposes including link capacity \cite{greedy algorithm,Giannakis}.
The case of RA with discrete-valued utilities requires a different
treatment.

\section{Numerical Examples}

In order to demonstrate the application of the SGPA RA algorithm in
a massive CA setting, an example case where $K=30$ UEs are served
via a single cell utilizing up to $M=50$ CCs, each with $N=100$
RBs is considered. Note that for $M=50$, the optimal RA problem formulation
consists of $KMN=150,000$ binary valued variables $\{\alpha_{k,m,n}\}$.
The normalized maximum transmission rate (link capacity) was considered
as the utility function \cite{greedy algorithm,Giannakis}, i.e., 

\begin{equation}
\phi_{k,m,n}=\frac{1}{N}\log_{2}\left(1+g_{k,m,n}\mathsf{SNR}_{k,m}\right),\forall k,m,n,\label{eq:capacity_utility}
\end{equation}

\noindent where the elements of $\{g_{k,m,n}\}$, representing channel
gains, are independent, identically distributed (i.i.d.) according
to an exponential p.d.f. of unit mean, and the elements of $\{\mathsf{SNR}_{k,m}\}$,
representing the average CC signal-to-noise-ratios per UE, are i.i.d.
according to a uniform p.d.f. over the interval $[-10,20]$ (in dB).
Note that for this choice of utility function and with $w_{k}=1$
for all $k$, the $\mathsf{WSU}$ is equal to the sum capacity of
the system.

In all cases, the SGPA algorithm was initialized as $\alpha_{k,m,n}^{(0)}=1/K$,
$\beta_{k,m}^{(0)}=1/M_{k}$, $\gamma_{m}^{(0)}=1/M_{0}$, for all
$k,m,n$, and was restricted to perform $20$ iterations irrespective
of the RA problem dimension. Since convergence has not been achieved
at this point in general, quantization of the final RA variables is
performed, obtained by setting, for each $k\in\mathcal{K},$ the largest
$M_{k}$ values of $\{\beta_{k,m}^{(20)}\}_{m\in\mathcal{M}}$ equal
to $1$ and zero, otherwise. The quantization of $\{\alpha_{k,m,n}^{(20)}\}$
and $\{\gamma_{m}^{(20)}\}$ is performed similarly.

For comparison purposes, a heuristic RA algorithm of comparable complexity
with the SGPA RA algorithm was considered, consisting of two steps.
First, by assuming $\alpha_{k,m,n}=1$ for all $k,m,n$, it solves
the relaxed RA problem w.r.t. $\{\beta_{k,m}\}$ and $\{\gamma_{m}\}$
after re-formulating it as a relaxed LP problem by introduction of
auxiliary variables \cite[Sec. 3.4]{Gotsis}. After quantization of
the LP solution with the same method as in the SGPA RA algorithm,
the $\{\alpha_{k,m,n}\}$ variables are determined by the rule of
(\ref{eq:greedy_assigments}) with $w_{k}\gamma_{m}\beta_{k,m}\phi_{k,m,n}$
in place of $w_{k}\phi_{k,m,n}$. 

\begin{figure}[t]
\noindent \centering{}\includegraphics[width=0.95\columnwidth]{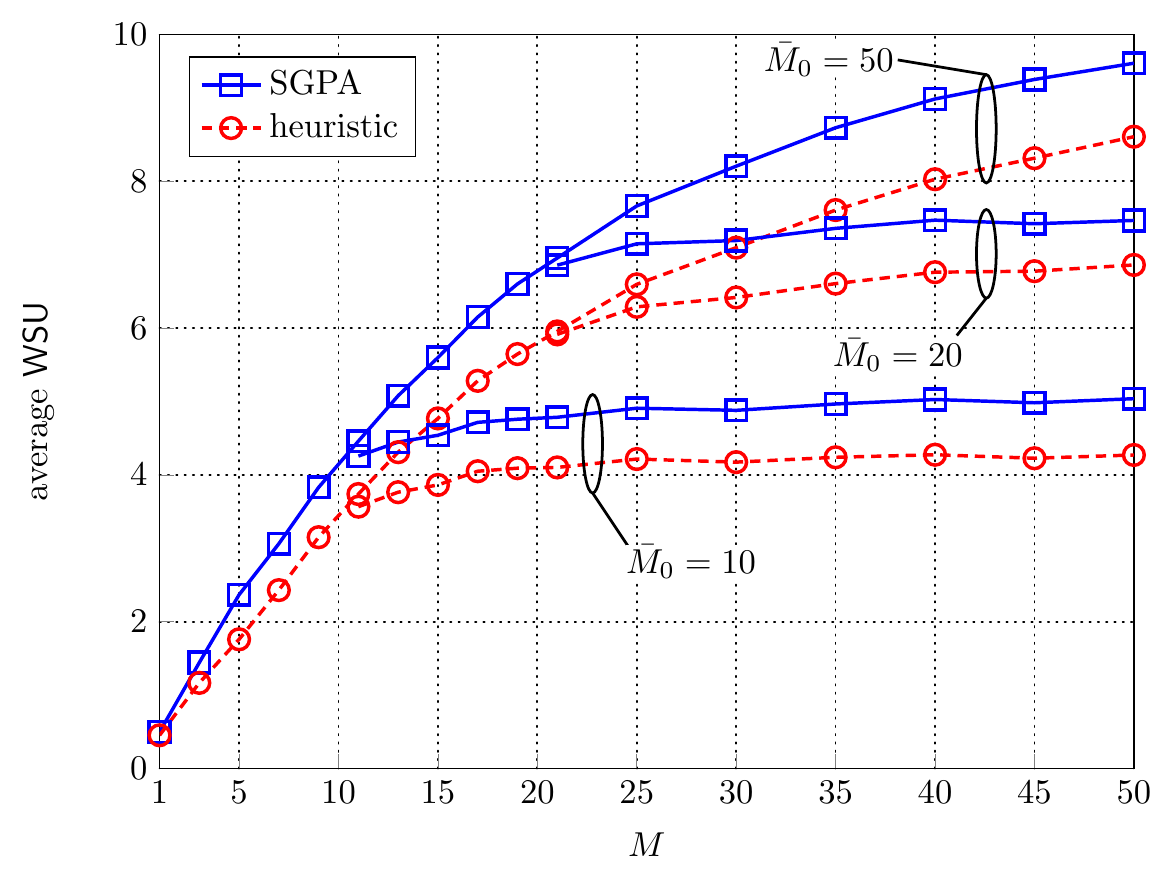}\caption{Average $\mathsf{WSU}$ performance of RA algorithms with varying
$M$.\label{fig:WSR_vs_M}}
\end{figure}

Fig. \ref{fig:WSR_vs_M} shows the $\mathsf{WSU}$ obtained by averaging
over independent realizations of utilities $\{\phi_{k,m,n}\}$ and
UE weights, with the latter uniformly distributed over the set $\{w_{k},k\in\mathcal{K}:w_{k}\geq0,\forall k,\sum_{k\in\mathcal{K}}w_{k}=1$\}.
The maximum number of CCs per UE was set to $M_{k}=2$ for all $k\in\mathcal{K}$,
whereas the maximum number of CCs used by the system was set to $M_{0}=\min\{M,\bar{M}_{0}\}$,
with $\bar{M}_{0}=10,20,50$. It can be seen that, in all cases, increasing
$M$ monotonically improves performance, since more resources are
available for RA purposes. However, when $M_{0}$ is limited to a
maximum value, consideration of $M>M_{0}$ provides only small gain,
since the total number of available RBs for RA purposes remains the
same. In comparison with the heuristic algorithm, the SGPA RA algorithm
provides better performance by approximately $10\%$ in all cases. 

\begin{figure}[tbh]
\noindent \centering{}\includegraphics[width=0.95\columnwidth]{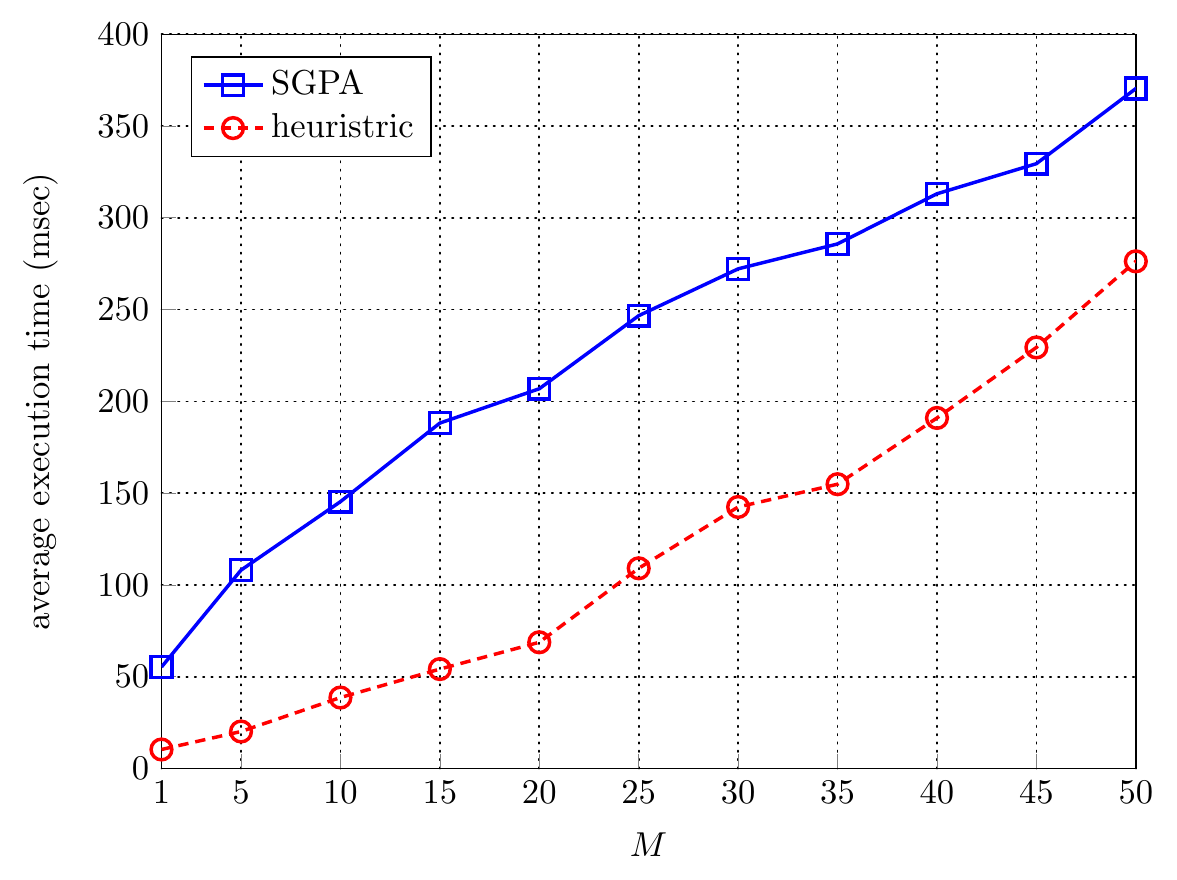}\caption{\label{fig:timings}Average execution time of RA algorithms with varying
$M$ ($\bar{M}_{0}=20)$.}
\end{figure}

Fig. \ref{fig:timings} depicts that average execution time of the
SGPA and heuristic algorithms for the case shown in Fig. \ref{fig:WSR_vs_M}
corresponding to $\bar{M}_{0}=20$ (results are similar for other
values of $\bar{M}_{0}$). A straightforward implementation in python
was considered for the SGPA algorithm, while the LP solver of \cite{ECOS}
was used for the heuristic algorithm. The measurements were performed
on an Intel i5 core operating on Linux. It can be seen that both algorithms
scale linearly with the problem dimension, which is essential for
their practical implementation in massive CA applications. For the
considered implementations, the SGPA algorithm is slower than the
heuristic, which can be viewed as the price to pay for the improved
performance. Note that the SGPA algorithm speed can be improved by
reducing the number of performed iterations, with a cost in average
$\mathsf{WSU}$ performance.

\begin{figure}[t]
\noindent \begin{centering}
\includegraphics[width=0.95\columnwidth]{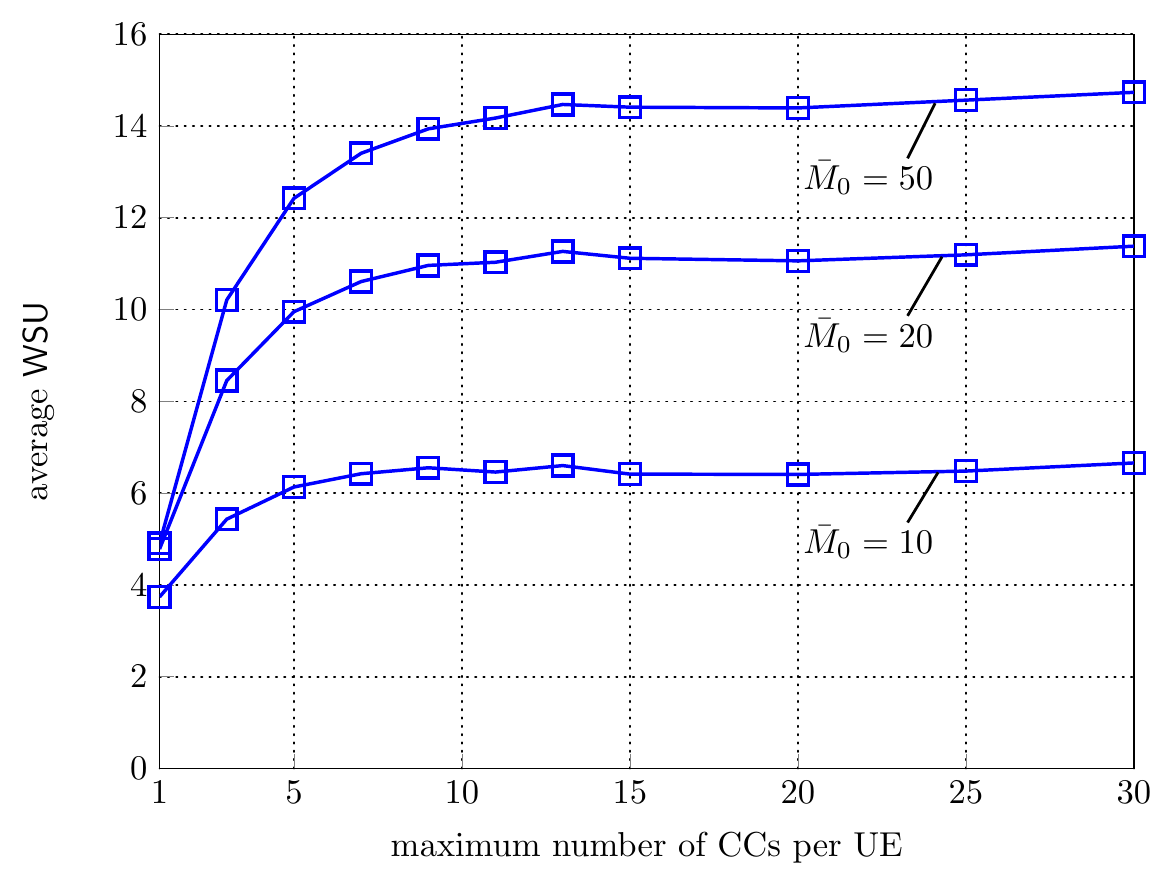}\caption{\label{fig:WSU_vs_Mk}Average $\mathsf{WSU}$ performance of the SGPA
algorithm with varying maximum number of CCs per UE ($M=50$).}
\par\end{centering}
\end{figure}

The effect of increasing the maximum CCs per UE is shown in Fig. \ref{fig:WSU_vs_Mk},
where the average $\mathsf{WSU}$ achieved via the SGPA algorithm
is depicted for the case $M=50$. All UEs were set to have the same
maximum number of CCs. As expected, increasing the maximum number
of CCs per UE increases performance. However, this increase is substantial
only up to a moderate number of CCs, above which, the ability of UEs
to transmit to more CCs offers marginal gains. This is an interesting
observation as it suggests that the cost associated with implementing
devices able to communicate with multiple CCs is unnecessary for leveraging
the benefits of massive CA.

\section{Conclusion}

An efficient iterative RA algorithm was proposed for massive CA scenarios.
It was shown that the algorithm obtains the optimal RA solution under
special cases of the RA problem, converges to binary-valued RA variables
in the general case under mild assumptions on the distribution of
utilities, and outperforms heuristic RA schemes.

\appendices{}

\section{Proof of Proposition \ref{prop:hardware arithmentic}}

For simplicity and without loss of generality, the case where $P_{0}=P$
will be considered, i.e., $r_{p}^{(i)}>0$ for all $p\in\mathcal{P}$.
The following result is critical as it shows that at any stage of
the iterations procedure before a fixed point is reached, one of the
sequences $x_{p}^{(i)}$ is monotonically increasing towards the limit
$1$.
\begin{lem}
\label{lem:basic lemma}Consider the RA algorithm initialization and
the ordering of $\{r_{p}^{(i)}\}$ as described in the statement of
Proposition \ref{prop:hardware arithmentic}. Assume that at iteration
$i-1,i\geq1$, the RA algorithm has not reached a fixed point and
let $\mathcal{L}^{(i-1)}\triangleq\{p\in\mathcal{P}:x_{p}^{(i-1)}=1\}$.
Let $p^{*}$ denote the sequence index such that $x_{p^{*}}^{(i-1)}<1$
with $r_{p^{*}}^{(i-1)}>r_{p}^{(i-1)}$ for all $p\neq p^{*},p\notin\mathcal{L}^{(i-1)}$.
It holds 

\begin{enumerate}
\item $x_{p}^{(j)}=1$ for all $j\geq i$ and $p\in\mathcal{L}^{(i-1)}$
with $r_{p}^{(i-1)}>r_{p^{*}}^{(i-1)}$, 
\item Sequence $x_{p^{*}}^{(j)},j\geq i,$ is strictly monotonically increasing
towards the limit value of $1$. 
\end{enumerate}
\end{lem}
\begin{IEEEproof}
First note that it must hold $|\mathcal{L}^{(i-1)}|<L$, where $|\mathcal{L}^{(i-1)}|$
is the cardinality of $\mathcal{L}^{(i-1)}$, since, otherwise, the
iterations would have reached the limit point where $x_{p}^{(j)}=1,p\in\mathcal{L}^{(i-1)}$
and $x_{p}^{(j)}=0,p\notin\mathcal{L}^{(i-1)}$, for all $j\geq i-1$.
In addition, note that the iteration scheme of (\ref{eq:general_update_rule})
guarantees that, with $x_{p}^{(0)}>0$, $x_{p}^{(i)}\in(0,1]$ for
all $i\geq0$ such that $|\mathcal{L}^{(i)}|<L$. Therefore, index
$p^{*}$ exists. Let $\mathcal{L}_{*}^{(i-1)}\triangleq\{p\in\mathcal{L}^{(i-1)}:r_{p}^{(i-1)}>r_{p^{*}}^{(i-1)}\}\subseteq\mathcal{L}^{(i-1)}$.
The normalization factor $\kappa^{(i)}$ of (\ref{eq:general_update_rule})
is found as the unique solution of (\ref{eq:sum constraint-1}). It
is easy to verify that (\ref{eq:sum constraint-1}) can only be satisfied
if $\kappa^{(i)}\leq\min\{r_{p}^{(i)}\}_{p\in\mathcal{L}_{*}^{(i-1)}}$.
This condition implies from (\ref{eq:general_update_rule}) that $x_{p}^{(i)}=1$,
for all $p\in\mathcal{L}_{*}^{(i-1)}$. By repeating this argument
for the next iterations, it follows that $x_{p}^{(j)}=1$,$j\geq i$,
$p\in\mathcal{L}_{*}^{(i)}$, thus proving the fist claim.

Towards proving the second claim, note that $\kappa^{(i)}$ should
satisfy either $\kappa^{(i)}\leq x_{p^{*}}^{(i-1)}r_{p^{*}}^{(i)}$
or $\kappa^{(i)}>x_{p^{*}}^{(i-1)}r_{p^{*}}^{(i)}$. In the former
case, $x_{p^{*}}^{(i)}=1,$ and by the previous argument, it follows
that this value remains fixed for all subsequent iterations. In the
latter case, $x_{p^{*}}^{(i)}=x_{p^{*}}^{(i-1)}r_{p^{*}}^{(i)}/\kappa^{(i)}$,
whose value can be lower bounded by finding an upper bound for $\kappa^{(i)}$.
Towards this end, it follows from (\ref{eq:sum constraint-1}) that 

\begin{align*}
L & =\sum_{p\in\mathcal{L}^{(i-1)}}\min\left\{ 1,\frac{r_{p}^{(i)}}{\kappa^{(i)}}\right\} +\sum_{p\notin\mathcal{L}^{(i-1)}}\frac{x_{p}^{(i-1)}r_{p}^{(i)}}{\kappa^{(i)}}\\
 & \leq|\mathcal{L}^{(i-1)}|+\frac{1}{\kappa^{(i)}}\sum_{p\notin\mathcal{L}^{(i-1)}}x_{p}^{(i-1)}r_{p}^{(i)}.
\end{align*}

\noindent Therefore, 

\begin{align}
\kappa^{(i)} & \leq\frac{\sum_{p\notin\mathcal{L}^{(i-1)}}x_{p}^{(i-1)}r_{p}^{(i)}}{L-|\mathcal{L}^{(i-1)}|}\nonumber \\
 & =\frac{r_{p^{*}}^{(i)}\left(x_{p^{*}}^{(i-1)}+\sum_{p\notin\mathcal{L}^{(i-1)}\setminus p^{*}}x_{p}^{(i-1)}\frac{r_{p}^{(i)}}{r_{p^{*}}^{(i)}}\right)}{L-|\mathcal{L}^{(i-1)}|}\nonumber \\
 & \stackrel{(a)}{<}\frac{r_{p^{*}}^{(i)}\left(x_{p^{*}}^{(i-1)}+\sum_{p\notin\mathcal{L}^{(i-1)}\setminus p^{*}}x_{p}^{(i-1)}\right)}{L-|\mathcal{L}^{(i-1)}|}\nonumber \\
 & \stackrel{(b)}{=}\frac{r_{p^{*}}^{(i)}\left(L-\sum_{p\in\mathcal{L}^{(i-1)}}x_{p}^{(i-1)}\right)}{L-|\mathcal{L}^{(i-1)}|}\nonumber \\
 & \stackrel{(c)}{=}r_{p^{*}}^{(i)},\label{eq:appendix}
\end{align}

\noindent where the assumption $r_{p^{*}}^{(i)}>r_{p}^{(i)}$ for
all $p\neq p^{*},p\notin\mathcal{L}^{(i-1)}$, was used in $(a)$,
$(b)$ follows from (\ref{eq:sum constraint-1}) evaluated at iteration
$i-1$ and $(c)$ follows since $x_{p}^{(i-1)}=1$ for $p\in\mathcal{L}^{(i-1)}$.
Therefore, $x_{p^{*}}^{(i)}>x_{p^{*}}^{(i-1)}$, i.e., $x_{p^{*}}^{(i)}$
is an increasing bounded sequence, which means that it has a limit
$\bar{x}_{p^{*}}$. It follows from (\ref{eq:general_update_rule})
that this limit must satisfy the condition

\begin{equation}
\bar{x}_{p^{*}}=\min\left\{ 1,\bar{x}_{p^{*}}\underset{i\rightarrow\infty}{\lim}\left(r_{p^{*}}^{(i)}/\kappa^{(i)}\right)\right\} .\label{eq:appendix 2}
\end{equation}

\noindent Noting from (\ref{eq:appendix}) that $r_{p^{*}}^{(i)}/\kappa^{(i)}>1$,
for all $i$, it follows that $\bar{x}_{p^{*}}=1$ is the only positive
value that can satisfy (\ref{eq:appendix 2}).
\end{IEEEproof}
The proof of Proposition \ref{prop:hardware arithmentic} now directly
follows by noting that at iteration $1$ there will exist an index
$p^{*}$ such that $x_{p^{*}}^{(i)},i\geq1,$ will be strictly monotonically
increasing towards the limit $1$, according to Lemma \ref{lem:basic lemma}.
Therefore, for any $\epsilon>0$, there exists an iteration index,
say, $i_{1}$, such that $x_{p^{*}}^{(i_{1})}\geq1-\epsilon$. When
$\epsilon$ equals the finite precision used in the implementation
of the algorithm, $x_{p^{*}}^{(i_{1})}$ will be set equal to $1$.
By Lemma \ref{lem:basic lemma}, $x_{p^{*}}^{(i)}=1$ for all $i\geq i_{1}$
and a new index $q^{*}$ will exist such that $x_{q^{*}}^{(i)}$ will
start to monotonically increase towards $1.$ It is easy to see that
this procedure repeats until all sequences of index $p$ such that
$r_{p}^{(i)}\geq r_{\pi(L)}^{(i)}$ will have achieved the value of
$1$.

\end{document}